\documentstyle[floats,aps,psfig]{revtex}
\begin{document}

\draft

\title{
Generalization properties of finite size polynomial Support Vector Machines 
}

\author{Sebastian Risau-Gusman and Mirta B. Gordon\\   
DRFMC/SPSMS CEA Grenoble, 17 av. des Martyrs\\  
38054 Grenoble cedex 09, France  
}  

\date{\today}
  
\maketitle  
  
\begin{center}
\begin{abstract}
  
\parbox{14cm}{
The learning properties of finite size polynomial 
Support Vector Machines are analyzed in the case 
of realizable classification tasks. The normalization 
of the high order features acts as a squeezing 
factor, introducing a strong anisotropy in the patterns 
distribution in feature space. As a function of the training 
set size, the corresponding generalization 
error presents a crossover, more or less abrupt depending on 
the distribution's anisotropy and on the task to be 
learned, between a fast-decreasing and 
a slowly decreasing regime. This behaviour corresponds to the stepwise 
decrease found by Dietrich et al.~[1] in the thermodynamic 
limit. The theoretical results are in excellent agreement 
with the numerical simulations. 
} 
\end{abstract}
\end{center}


\section{Introduction}

In the last decade, the typical properties of neural networks 
that learn classification tasks from a set of examples have been 
analyzed using the approach of Statistical Mechanics. In the 
general setting, the value of a binary output neuron represents 
whether the input vector, describing a particular pattern, 
belongs or not to the class to be recognized. Manuscript character 
recognition and medical diagnosis are examples of such 
classification problems. The process of inferring the rule 
underlying the input-output mapping given a set of examples 
is called learning. The aim is to predict correctly the class 
of novel data, i.e. to generalize.

In the simplest neural network, the perceptron, the inputs 
are directly connected to a single output neuron. The output 
state is given by the sign of the weighted sum of the inputs. Then, 
learning amounts to determine the weights of the connexions in 
order to obtain the correct outputs to the training examples. 
Considering the weights as the components of a vector, 
the network classifies the input vectors according to whether 
their projections onto the weight vector are positive or negative. 
Thus, patterns of different classes are separated by the hyperplane 
orthogonal to the weight vector. Beyond these linear separations, 
two different learning schemes have been suggested. Either the input 
vectors are mapped by linear hidden units to so called internal 
representations that must be linearly separable by the output 
neuron, or a more powerful output unit is defined, 
able to perform more complicated functions than just the weighted 
sum of its inputs. 

The first solution is implemented using 
feedforward layered neural networks. The classification of the 
internal representations, performed by the output neuron, 
corresponds in general to a complicated separation surface 
in input space. However, the relation between the number of 
hidden units of a network and the class of rules it can infer is 
still an open problem. In practice, the number of hidden 
neurons is either guessed or determined through constructive 
heuristics.  

A solution that uses a more complex output unit, the 
Support Vector Machine (SVM)~\cite{Vapnik}, has been 
recently proposed. The input patterns are transformed into
high dimensional feature vectors whose components 
may include the original input together with specific 
functions of its coordinates selected a priori, with the aim  
that the learning set be {\it linearly separable} 
in feature space. In that case the learning problem 
is reduced to that of training a simple perceptron. For example, 
if the feature space includes all the pairwise products of 
the input vector, the SVM may implement any classification rule 
corresponding to a quadratic separating surface in input space. 
Higher order polynomial SVMs and other types of SVMs 
may be defined by introducing the corresponding features.
A big advantage is that learning a linearly separable rule 
is a convex optimization problem. The difficulties of having 
many local minima, that hinder the process of training 
multilayered neural networks, are thus circumvented.
Once the adequate feature space is defined, the 
SVM selects the particular hyperplane called Maximal Margin 
(or Maximal Stability) Hyperplane (MMH), which lies at the largest 
distance to its closest patterns in the training set. These patterns are 
called Support Vectors (SV). The MMH solution has 
interesting properties~\cite{OKKN}. In particular, 
the fraction of learning patterns that belong to the 
SVs provides an upper bound~\cite{Vapnik} to the 
generalization error, that is, to the probability of 
incorrectly classifying a new input. It has been 
shown~\cite{OK} that the perceptron 
weights are a linear combination of the SVs, an interesting 
property in high dimensional feature spaces, as their 
number is bounded. 

A perceptron can learn with very high probability any set of 
examples, regardless of the underlying classification rule, 
provided that their number does not exceed twice its input 
space dimension~\cite{Cover}. However, this simple rote 
learning does not capture the rule underlying the 
classification. As it may arise that the feature space 
dimension of the SVM is comparable to, or even larger 
than, the number of available training patterns, we would 
expect that SVMs have a poor generalization performance. 
Surprisingly, this seems not to be the case in the 
applications~\cite{CV}. 

Two theoretical papers~\cite{DOS,BG} have recently addressed this 
interesting question. They determined the typical properties 
of a family of polynomial SVMs in the limit of large dimensional 
spaces, reaching completely different results in spite of the 
seemingly innocuous differences between the models. Both papers 
consider polynomial SVMs in which the input vectors 
${\bf x} \in \mbox{I\hspace{-0.18cm}R}^N$ are 
mapped onto quadratic features $\bbox \Phi$. More 
precisely, the {\it normalized} mapping ${\bbox \Phi}_n ({\bf x})=({\bf x}, 
x_1{\bf x}/\sqrt{N}, x_2{\bf x}/\sqrt{N}, \cdots, 
x_N{\bf x}/\sqrt{N})$ has been considered in~\cite{DOS}. 
The {\it non-normalized} mapping 
${\bbox \Phi}_{nn} ({\bf x})=({\bf x},x_1{\bf x}, 
x_2{\bf x}, \cdots, x_{k}{\bf x})$ has been studied in~\cite{BG} 
as a function of $k$, the  number of quadratic features. For $k=N$ 
the dimension of both feature spaces is the same, corresponding to 
a {\it linear subspace} of dimension $N$, and a {\it quadratic subspace} 
of dimension $N^2$. The mappings only differ in the distributions 
of the quadratic components in feature space. Due to the 
normalization, those of ${\bbox \Phi}_n$ 
are squeezed by a {\it normalizing factor} $a=1/\sqrt{N}$ 
with respect to those of ${\bbox \Phi}_{nn}$. In the case of 
learning a linearly separable rule with 
the non-normalized mapping ${\bbox \Phi}_{nn}$, the generalization 
error at any given learning set size increases dramatically 
with the number $k$ of quadratic features included~\cite{BG}.
On the contrary, in the case of mapping ${\bbox \Phi}_{n}$, 
the generalization error exhibits an interesting stepwise 
decrease, also found within the Gibbs learning 
paradigm in a quadratic feature space~\cite{YO}. 
If the number of training patterns scales with $N$, 
the dimension of the linear subspace, it decreases 
up to an asymptotic lower bound. If the number 
of examples scales proportionally to $N^2$, it 
vanishes asymptotically. In particular, 
if the rule to be inferred is linearly separable in the 
input space, learning in the feature space with the mapping 
${\bbox \Phi}_{n}$ is harmless, as the decrease of the 
generalization error with the number of training patterns 
presents a slight slow-down with respect to that of a 
simple perceptron learning in input space. 

As this stepwise learning is {\it exclusively} related 
to the fact that the normalizing factor of the quadratic 
features vanishes in the thermodynamic limit $N \rightarrow 
\infty$, in the present paper we determine the influence of 
the normalizing factor on the typical generalization 
performance of finite size SVMs. To this end, we introduce two 
parameters, $\sigma$ and $\Delta$, caracterizing the mapping of 
the $N$-dimensional input patterns onto the feature 
space. The {\it variance} $\sigma$ reflects the 
width of the high-order features distribution and is
related to the normalizing factor $a$. The {\it inflation 
factor} $\Delta$ accounts for the proportion of quadratic 
features with respect to the input space 
dimension $N$. Actual quadratic SVMs are caracterized 
by different values of $\Delta$ and $\sigma$, 
depending on $N$ and $a$. Keeping $\sigma$ and 
$\Delta$ fixed in the thermodynamic limit allows 
us to determine the typical properties 
of actual SVMs, which have finite compressing 
factors and inflation ratios. 

In fact, the behaviour of the SVMs is the 
same as that of a simple perceptron learning a 
training set with patterns drawn from a highly 
anisotropic probability distribution, such that 
a {\it  macroscopic} fraction of components have a 
different variance from the others. Not surprisingly, 
we find that the asymptotic behaviour corresponding 
to both the small and large training set size limits, 
is the same as the one of the perceptron's MMH. 
Only the prefactors depend on the mapping used by the SVM. 

As expected, the stepwise learning obtained with 
the normalized mapping in the thermodynamic limit 
becomes a crossover. Upon increasing the 
number of training patterns, the generalization 
error first present an abrupt decrease, that corresponds 
to learning the weight components in the linear subspace, 
followed by a slower decrease corresponding to the learning 
of the quadratic components. The steepness of the crossover 
not only depends on $\Delta$ and $\sigma$, but also on 
the task to be learned. The agreement between our analytic 
results and numerical simulations is excellent. 

The paper is organized as follows: in section \ref{sec:model} 
we introduce the model and the main steps of the Statistical 
Mechanics calculation. Numerical simulation results are compared  
to the corresponding theoretical predictions
in section \ref{sec:results}. The two 
regimes of the generalization error and the 
asymptotic behaviours are discussed in section 
\ref{sec.discussion}. The conclusion 
is left to section \ref{sec:conclusion}. 
  
\section{The model} 
\label{sec:model} 

We consider the problem of learning a binary classification 
task from examples with a SVM in polynomial feature spaces. 
The learning set contains $M$ patterns $({\bf x}^\mu, 
\tau^\mu)$ ($\mu=1,\cdots,M$) where ${\bf x}^\mu$ 
is an input vector in the $N$-dimensional input space, 
and $\tau^\mu \in \{-1,1\}$ 
is its class. We assume that the components $x_i^\mu$ ($i=1, 
\cdots, N$) are independent identically distributed (i.i.d.) 
random variables drawn from gaussian distributions 
having zero-mean and unit variance:

\begin{equation}
\label{eq.pdex}  
P({\bf x}) = \prod_{i=1}^{N} \frac{1}{\sqrt{2 \pi}} \exp \left( - \frac{ x_i^2} {2} \right).  
\end{equation} 

\noindent In the following we concentrate on quadratic feature 
spaces, although our conclusions are more general, and may be 
applied to higher order polynomial SVMs, as discussed in section 
\ref{sec.discussion}. The mappings 
${\bbox \Phi}_{nn} ({\bf x})=({\bf x},x_1{\bf x}, x_2{\bf x}, 
\cdots, x_N{\bf x})$ and  ${\bbox \Phi}_n ({\bf x})=({\bf x}, 
x_1{\bf x}/\sqrt{N},  x_2{\bf x}/\sqrt{N}, \cdots, 
x_N{\bf x}/\sqrt{N})$ are particular instances of mappings 
of the form ${\bbox \Phi} ({\bf x})= (\phi_1, 
\phi_2, \cdots, \phi_N, \phi_{11}, 
\phi_{12},\cdots, \phi_{NN})$ where $\phi_i=x_i$, and 
$\phi_{ij}=a \, x_i x_j$, where $a$ is the normalizing 
factor of the quadratic components: $a=1$ 
for mapping ${\bbox \Phi}_{nn}$ and $a=1/\sqrt{N}$ for 
${\bbox \Phi}_{n}$. The patterns probability 
distribution in feature-space is:

\begin{equation}
\label{eq.pdephi}
P\left({\mbox{$\bbox \Phi$}}\right) = \int \, \prod_{i=1}^{N} 
\frac{dx_i}{\sqrt{2 \pi}} \exp \left( - \frac{ x_i^2} {2} \right) 
\delta(\phi_i - x_i)
\prod_{j=1}^{N} \delta \left(\phi_{ij}-a \, x_i x_j \right). 
\end{equation}
 
\noindent Clearly, the components of ${\bbox \Phi}$ are not independent 
random variables. For example, a number $O(N^3)$ of triplets of the 
form $\phi_{ij} \phi_{jk} \phi_{ki}$ have positive correlations. 
These contribute to the third order moments, which should vanish 
if the features were gaussian. Moreover, the fourth order connected 
correlations~\cite{Monasson} do not vanish in the thermodynamic 
limit. Nevertheless, in the following we will neglect these 
and higher order connected moments. This approximation, used 
in~\cite{BG} and implicit in~\cite{DOS}, is equivalent to 
assuming that all the components in feature space are 
independent gaussian variables. Then, the only difference 
between the mappings $\bbox \Phi_{n}$ and $\bbox \Phi_{nn}$ 
lies in the variance of the quadratic 
components distribution. The results obtained using this 
simplification are in excellent agreement with the 
numerical tests described in the next section.

Since, due to the symmetry of the 
transformation, only $N(N+1)/2$ among the $N^2$ quadratic 
features are different, hereafter we restrict the 
feature space and only consider the non redundant 
components, that we denote 
${\bbox \xi} =({\bbox \xi}_u,{\bbox \xi}_\sigma)$. 
Its first $N$ components 
${\bbox \xi}_u = (\xi_1, \cdots, \xi_N)$ hereafter called 
{\it u-components}, represent the input pattern of unit variance, 
lying in the linear subspace. 
The remaining components ${\bbox \xi}_\sigma= (\xi_{N+1}, 
\cdots, \xi_{\tilde N})$ stand for the {\it non redundant} 
quadratic features, of variance $\sigma$, hereafter 
called $\sigma$-{\it components}. $\tilde N$ is 
the dimension $\tilde N = N(1+\Delta)$ of the 
restricted feature space, where the 
{\it inflation ratio} $\Delta$ is 
the relative number of non-redundant quadratic features 
per input space dimension. The quadratic mapping has 
$\Delta=(N+1)/2$. 

According to the preceding 
discussion, we assume that learning $N$-dimensional 
patterns selected with the isotropic distribution 
(\ref{eq.pdex}) with a quadratic SVM is equivalent 
to learning the MMH with a simple perceptron in an 
$\tilde N$-dimensional space where the patterns are 
drawn using the following 
anisotropic distribution, 

\begin{equation}
\label{eq.pdexi}
P\left(\mbox{$\bbox \xi$}\right) = \prod_{i=1}^{N} 
\frac{1}{\sqrt{2 \pi}} \exp \left( - \frac{ \xi_i^2} {2} \right) \;
\prod_{j=N+1}^{\tilde N} \frac{1}{\sqrt{2 \pi \sigma^2}} \exp \left( - \frac{ \xi_j^2} {2 \sigma^2} \right). 
\end{equation} 

\noindent The second moment of the {\it u}-features 
is $\langle {\bbox \xi}_u^2 \rangle = N$ and that of the 
$\sigma$-features is $\langle {\bbox \xi}_\sigma^2 \rangle = N 
\Delta \sigma^2$. If $\sigma^2 \Delta=1$, we get 
$\langle {\bbox \xi}_\sigma^2 \rangle = \langle 
{\bbox \xi}_u^2 \rangle$, which is the relation satisfied 
by the normalized mapping considered in~\cite{DOS}. The 
non-normalized mapping corresponds to $\sigma^2 \Delta=N$. 
In the following, instead of selecting either of these possibilities 
a priori, we consider $\Delta$ and $\sigma$ as independent 
parameters, that are kept constant when taking the thermodynamic 
limit. 

Since the rules to be inferred are assumed to be linear 
separations in feature space, we represent them by the 
weights ${\bf w}^*=(w_1^*,w_2^*,\cdots,w_{\tilde N}^*)$ 
of a {\it teacher perceptron}, so that the class 
of the patterns is $\tau ={\rm sign}(\mbox{$\bbox \xi$} 
\cdot {\bf w}^*)$. Without 
any loss of generality we consider normalized teachers: 
${\bf w}^* \cdot {\bf w}^* = \tilde N$. The training set 
in feature space is then ${\mathcal L}_M=
\{ ({\bbox \xi}^\mu, \tau^\mu) \}_{\mu=1,\cdots,M}$. 

In the following we study the typical properties of polynomial 
SVMs learning realizable classification tasks, using the tools 
of Statistical Mechanics. If ${\bf w}=(w_1, \cdots, w_{\tilde N})$ 
is the student perceptron weight vector, $\gamma^\mu = 
\tau^\mu {\bbox \xi}^\mu \cdot \bf w / \sqrt{{\bf w} \cdot {\bf w}}$ 
is the {\it stability} of pattern $\mu$ in feature space. 
The pertinent cost function is :  

\begin{equation} 
\label{eq.cost} 
E({\bf w}, \kappa; {\mathcal L}_M)=\sum_{\mu=1}^M \Theta(\kappa-\gamma^\mu).
\end{equation}  

\noindent $\kappa$, the smallest allowed distance 
between the hyperplane and the training patterns, is 
called the margin. The MMH corresponds to the weights 
with vanishing cost (\ref{eq.cost}) that maximize $\kappa$. 

The typical properties of cost (\ref{eq.cost}) in the case 
of isotropic pattern distributions have been exhaustively 
studied~\cite{OKKN,GG}. The case of a single anisotropy axis 
has also been investigated~\cite{MBS}. Here we study the 
case of the anisotropic distribution (\ref{eq.pdexi}), 
where a {\it macroscopic} fraction of components have 
different variance from the others, which is  pertinent for 
understanding the properties of the SVM.

Considering the cost (\ref{eq.cost}) as an energy, the partition 
function at temperature $1/\beta$ writes

\begin{equation}
\label{eq.partition}
Z(\kappa, \beta; {\mathcal L}_M)=\int \exp [-\beta E({\bf w}, \kappa; {\mathcal L}_M) ] \; p({\bf w}) \, d{\bf w}.
\end{equation}

\noindent Without any loss of generality, we assume that the a 
priori distribution of the student weights is uniform over the 
hypersphere of radius $\tilde N^{1/2}$, i.e. 
$p({\bf w})=\delta({\bf w} \cdot {\bf w} - \tilde N)$, 
meaning that the student weights are normalized in feature 
space. In the limit $\beta \rightarrow \infty$, the 
corresponding free energy $f(\kappa, \beta; {\mathcal L}_M) 
= - (1/\beta N) \ln Z(\kappa, \beta; {\mathcal L}_M)$ is 
dominated by the weights that minimize the cost (\ref{eq.cost}).

The typical properties of the MMH are obtained by looking for 
the largest value of $\kappa$ for which the quenched average 
of the free energy over the patterns distribution, in the zero 
temperature limit $\beta \rightarrow \infty$, vanishes. This 
average is calculated by the replica method, using the identity

\begin{equation}
f(\kappa, \beta)=-\frac{1}{N \beta} \, \overline{\ln Z(\kappa,\beta;{\mathcal L}_M)} = -\frac{1}{N \beta} \, \lim_{n \rightarrow 0} 
\frac{\ln \overline{Z^n(\kappa,\beta;{\mathcal L}_M)} }{n},
\end{equation}

\noindent where the overline represents the average over 
${\mathcal L}_M$, composed of patterns selected according 
to (\ref{eq.pdexi}). 

We obtain the typical properties of the MMH corresponding to 
given values of $\Delta$ and  $\sigma$ by taking 
the thermodynamic limit $N \rightarrow \infty$, 
$M \rightarrow \infty$, with $\alpha \equiv M/N$, 
$\Delta$ and $\sigma$ constant. Notice that the relation 
between the number 
of training examples and the {\it feature} space dimension, 
$\tilde \alpha \equiv M/\tilde N = \alpha/(1+\Delta)$, is 
finite. Thus, not only are we able to study the dependence 
of the learning properties as a function of the training 
set size as usual, but also of the inflation factor that 
characterizes the SVM, as well as of the variance of the 
quadratic components. As we only consider realizable rules, 
i.e. classification tasks that are linearly separable in 
feature space, the energy (\ref{eq.cost}) is a convex 
function of the weights ${\bf w}$, and replica symmetry 
holds.

For any $\kappa < \kappa_{max}$, there are a macroscopic 
number of weights that minimize the cost function 
(\ref{eq.cost}). In particular, in the case of $\kappa=0$, 
the cost is the number of training errors, 
and is minimized by any weight vector that classifies 
correctly the training set. The typical properties of such 
solution, called Gibbs learning, may be expressed in terms 
of several order parameters~\cite{RG}. Among them, $q_u^{ab}= 
\sum_{i=1}^N \langle \overline{w_i^a w_i^b} \rangle/\tilde N$, 
$q_\sigma^{ab}= \sum_{i=N+1}^{\tilde N}  \langle
\overline{w_i^a w_i^b} \rangle/\tilde N$ and $Q^a = 
\sum_{i=N+1}^{\tilde N}\langle \overline{w_i^a w_i^a} 
\rangle/\tilde N$, where $a \neq b$ are replica indices and 
$\langle \cdots \rangle$ stands for the usual 
thermodynamic average (with Boltzmann factor corresponding 
to the partition function (\ref{eq.partition})). $q_u^{ab}$ 
and $q_\sigma^{ab}$ represent the overlaps between different 
solutions in the {\it u}- and the $\sigma$- subspaces respectively. 
$\tilde N \, Q^a$ is the typical norm of the 
$\sigma$-components of replica $a$. 
Because of replica symmetry we have $Q^a$=$Q^b=Q$, 
$q_\sigma^{ab}=q_\sigma$ and $q_u^{ab}=q_u$ for all $a$, 
$b$. Upon increasing $\kappa$, the volume of the error-free 
solutions in weight space shrinks, and vanishes when 
$\kappa$ is maximized. Correspondingly, $q_u \rightarrow 1-Q$ 
and $q_\sigma \rightarrow Q$, with $x \equiv lim_{\kappa 
\rightarrow \kappa_{max}} (1-q_u/(1-Q))/(1-q_\sigma/Q)$ 
finite. In the limit of $\kappa \rightarrow \kappa_{max}$, 
the properties of the MMH may be expressed in 
terms of $x$, $\kappa_{max}$ and the following order parameters,

\begin{eqnarray} 
\label{eq.Q}
Q &=& \frac{1}{\tilde N} \sum_{i = N+1}^{\tilde N} \overline{\langle w_i^2 \rangle},
\\
R_u &=& \frac{1}{\sqrt{(1-Q)(1-Q^*)}} \frac{1}{\tilde N} \sum_{i=1}^N \overline{\langle w_i w^*_i \rangle}, 
\\
\label{eq.Rsigma}
R_\sigma &=& \frac{1}{\sqrt{Q \, Q^*}} \frac{1}{\tilde N} \sum_{i=N+1}^{\tilde N} \overline{\langle w_i w^*_i \rangle},
\end{eqnarray}

\noindent where $Q^*= \sum_{i = N+1}^{\tilde N} (w^*_i)^2/\tilde N$ 
is the teacher's squared weight vector in the 
$\sigma$-subspace. $Q$ is the corresponding typical value 
for the student. $R_u$ and $R_\sigma$ are proportional to 
the overlaps between the student and the teacher weights in 
the {\it u}- and the $\sigma$- subspaces respectively. 
The factors in the denominators arise because the weights are 
not normalized in each subspace.

The saddle point equations corresponding to the extremum 
of the free energy for the MMH are

\begin{eqnarray}
\label{eq.sp1}
2 \frac{\alpha}{\Delta} \Delta_\sigma I_1 &=& (1-R_\sigma^2) \,
\frac{(x + \Delta_\sigma)^2}{1+\Delta_\sigma}, 
\\
\label{eq.sp2}
2 \frac{\alpha}{\Delta}  I_2 
&=& \sqrt{\frac{1+\Delta^*_\sigma}{\Delta^*_\sigma}} \, R_\sigma \,
\frac{x + \Delta_\sigma}{\sqrt{\Delta_\sigma (1+\Delta_\sigma)}}, 
\\
\label{eq.sp3}
2 \frac{\alpha}{\Delta} Q (1-\sigma^2) I_3 &=& 
\left(1- x \frac{1-R_\sigma^2}{1-R_u^2}\right) \, 
\frac{x + \Delta_\sigma}{1+\Delta_\sigma},
\\
\label{eq.sp4}  
\frac{R_\sigma^2}{1-R_\sigma^2}  &=& \frac{\Delta^*_\sigma}{\Delta} \frac{R^2_u}{1-R^2_u},
\\
\label{eq.sp5}
\frac{R_u}{R_\sigma} &=& x \frac{\Delta}{\sqrt{\Delta_\sigma \Delta^*_\sigma}}.
\end{eqnarray}

\noindent where $\Delta_\sigma \equiv \sigma^2 Q/(1-Q)$ 
and $\Delta^*_\sigma \equiv \sigma^2 Q^*/(1-Q^*)$. The 
integrals in the left hand side of equations 
(\ref{eq.sp1}-\ref{eq.sp3}) are

\begin{eqnarray}
I_1 &=& \int_{-\tilde \kappa}^\infty Dt \, (t+{\tilde \kappa})^2 \, 
H\left(\frac{t R}{\sqrt{1-R^2}}\right),
\\
I_2 &=& \frac{1}{\sqrt{2 \pi}} \left[\frac{\sqrt{1-R^2} 
\exp(-{\tilde \kappa^2}/(2 (1-R^2)))} {\sqrt{2 \pi}} + 
{\tilde \kappa} H\left(\frac{-\tilde \kappa}{\sqrt{1-R^2}}\right)\right],
\\
I_3 &=& \int_{-\tilde \kappa}^\infty Dt \, \tilde \kappa \, (t+{\tilde \kappa}) \, H\left(\frac{t R}{\sqrt{1-R^2}}\right),
\end{eqnarray}

\noindent with $Dt \equiv dt \, \exp{(-t^2/2)}/\sqrt{2 \pi}$, 
$H(x)=\int_x^\infty Dt$, and 

\begin{eqnarray}
\tilde \kappa &=& \frac{\kappa_{max}}{\sqrt{(1-Q) (1+\Delta_\sigma)}},
\\
\label{eq.R}
R &=& \frac{R_u + \sqrt{\Delta_\sigma \Delta^*_\sigma} R_\sigma} 
{\sqrt{(1+\Delta_\sigma) (1+\Delta^*_\sigma)}}.
\end{eqnarray}

\noindent The value of $R$ determines the generalization 
error through $\epsilon_g=(1/\pi) \arccos(R)$.

After solving the above equations for $Q$, $R_u$, $R_\sigma$, 
$x$ and $\tilde \kappa$, it is straightforward to determine $\rho_{SV}$, 
the fraction of training patterns that belong to the subset of SV~\cite{O,GG,BG}: 

\begin{equation}
\rho_{SV} = 2 \int_{-\infty}^{\tilde \kappa} H\left(-tR/\sqrt{1-R^2}\right) Dt.
\end{equation}

In summary of this section, instead of considering 
a particular scaling of the fraction of high 
order features components and their 
normalization with $N$,  we analyzed the more 
general case where these quantities are kept as 
free parameters. We determined 
the saddle point equations that define the typical 
properties of the corresponding SVM. This approach 
allows us to consider several learning scenarios, 
and more interestingly, to study the crossover 
between the different generalization regimes.

\section{Results}
\label{sec:results}

\begin{figure}
\centerline{\psfig{figure=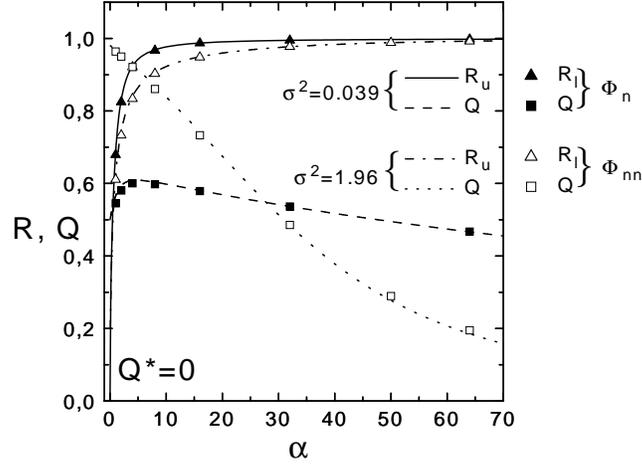,height= 7 cm}}
\caption{Order parameters of SVMs for purely linear 
teacher rules, $Q^*=0$. Symbols are experimental 
results for input space dimension $N=50$, corresponding 
to the two kinds of quadratic mappings, $\Phi_n$ 
with $a=1/\sqrt{N}$ (full symbols) and $\Phi_{nn}$ with 
normalizing factor $a=1$ (open symbols) respectively. 
Error bars are smaller than the symbols. The 
lines are solutions of equations ({\protect {\ref{eq.sp1}-\ref{eq.sp5}}}), 
for $\Delta=(N+1)/2$ and $\sigma^2=N a^2/\Delta$ with $N=50$, 
and $a$ corresponding to each mapping.}
\label{fig.Q*0}
\end{figure}

\begin{figure}
\centerline{\psfig{figure=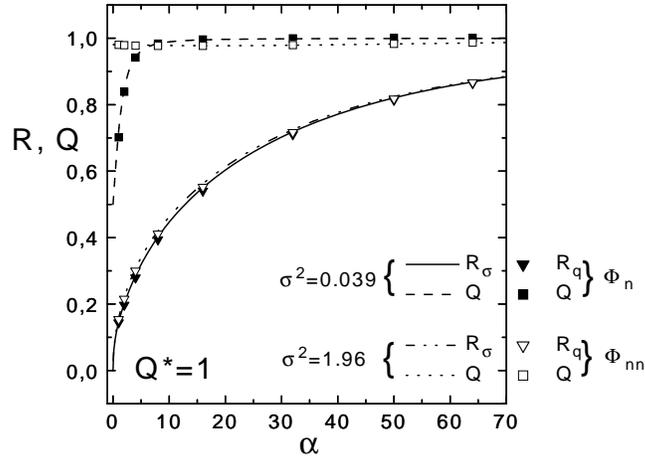,height= 7 cm}}
\caption{Order parameters of SVMs for purely quadratic 
teacher rules, $Q^*=1$. Definitions are the same 
as in figure {\protect {\ref{fig.Q*0}}}.}
\label{fig.Q*1}
\end{figure}

\begin{figure}
\centerline{\psfig{figure=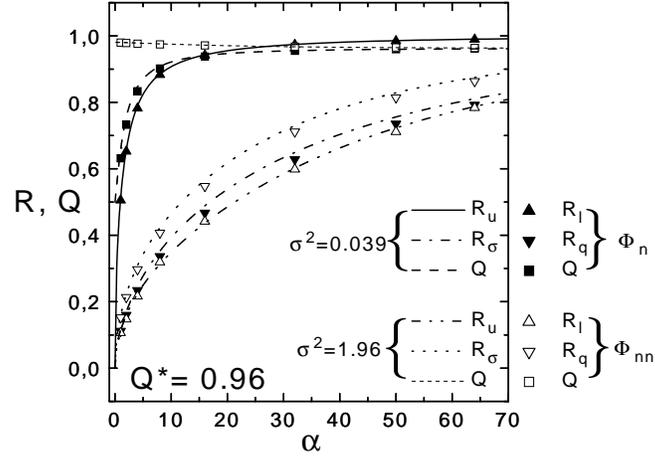,height= 7 cm}}
\caption{Order parameters of SVMs for isotropic 
teacher rules, $Q^*_{iso}=\Delta/(1+\Delta)$. Definitions are the same 
as in figure {\protect {\ref{fig.Q*0}}}.}
\label{fig.Q*51/53}
\end{figure}

\begin{figure}
\centerline{\psfig{figure=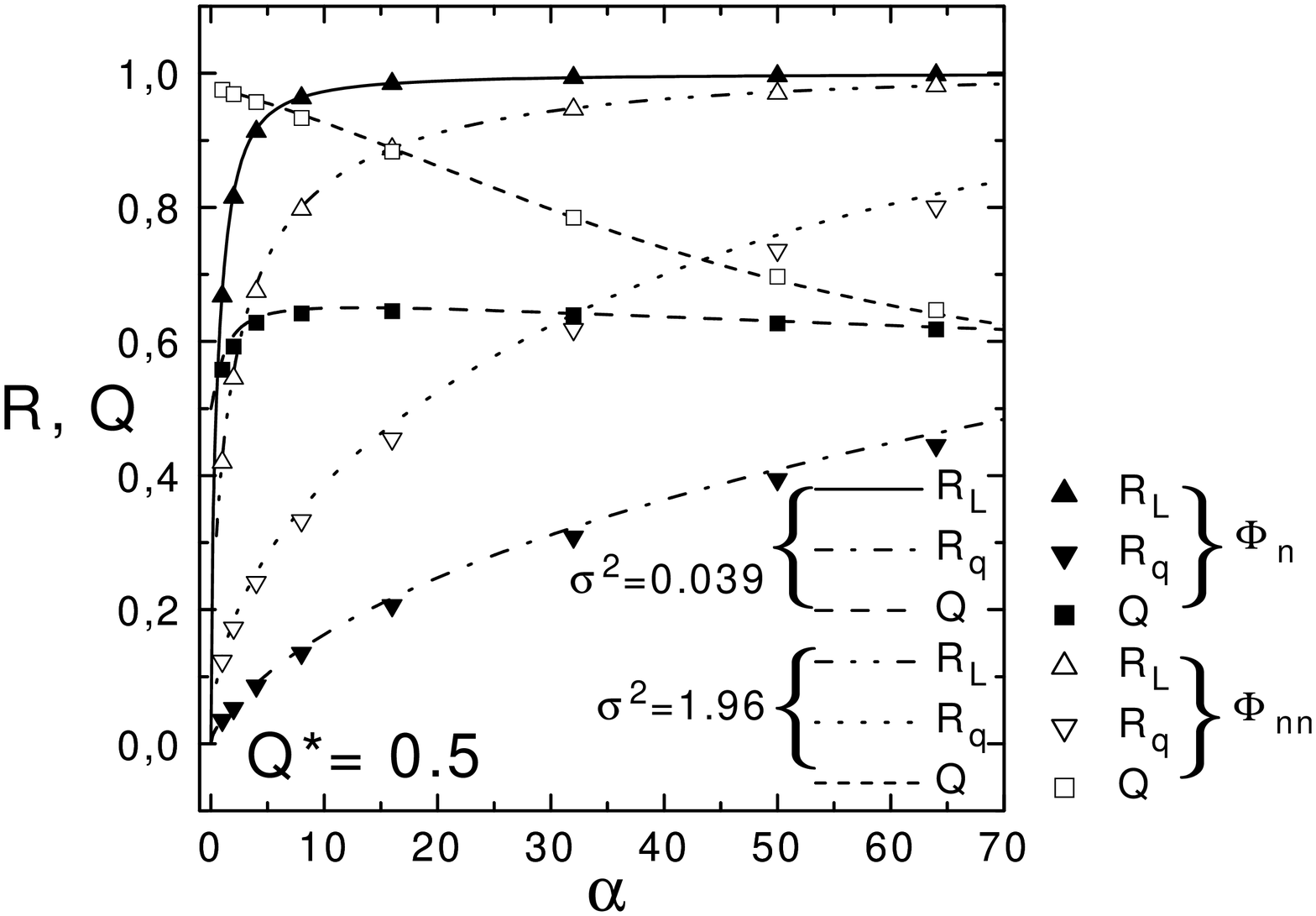,height= 7 cm}}
\caption{Order parameters of SVMs for a general 
teacher rule, $Q^*=0.5$. Definitions are the same 
as in figure {\protect {\ref{fig.Q*0}}}.}
\label{fig.Q*05}
\end{figure}

We describe first the experimental data, obtained 
with quadratic SVMs, using both mappings, ${\bbox \Phi}_{nn}$ and 
${\bbox \Phi}_n$, which have normalizing factors $a=1$ and 
$a=1/\sqrt{N}$ respectively, where $N$ is the 
input space dimension.  The $M= \alpha N$ random input 
examples of each training set 
were selected with probability (\ref{eq.pdex}) and 
labelled by teachers of normalized weights ${\bbox w}^* \equiv 
({\bbox w}^*_l,{\bbox w}^*_q)$ drawn at random. 
${\bbox w}^*_l$ are the $N$ components in the linear 
subspace and ${\bbox w}^*_q$ are the $N^2$ components 
in the quadratic subspace. Notice that, 
because of the symmetry of the mappings, teachers having 
the same value of the symmetrized weights in the 
quadratic subspace, $(w^*_{q,ij}+w^*_{q,ji})/2$, are 
all equivalent. The teachers are characterized by the proportion of 
(squared) weight components in the quadratic subspace, 
$Q^*={\bbox w}^*_q \cdot {\bbox w}^*_q/{\bbox w}^* \cdot {\bbox w}^*$. 
In particular, $Q^*=0$ and $Q^*=1$ correspond to a purely 
linear and a purely quadratic teacher respectively. 

The experimental student weights ${\bbox w} \equiv 
({\bbox w}_l,{\bbox w}_q)$ were obtained by 
solving numerically the dual problem~\cite{CV,V}, using the 
Quadratic Optimizer for Pattern Recognition program~\cite{AS}, 
that we adapted to the case without threshold treated in this paper. 
We determined $Q$, and the overlaps $R_l$ and $R_q$ 
in the linear and the quadratic subspaces, respectively. 
For each value of $M$, averages were performed over 
a large enough number of different teachers and training 
sets to get the precision shown in the figures. 

Experiments were carried out for 
$N=50$. The corresponding feature space dimension 
is $N(N+1)=2550$. The restricted feature space 
considered in our model is composed of the $N$ (linear) 
input components, which define the {\it u}-subspace 
of the feature space, and the $N \Delta$ 
non redundant quadratic components of the 
$\sigma$-subspace. For the sake of comparison 
with the theoretical results determined in the 
thermodynamic limit, we caracterize the actual 
SVM by its (finite size) inflation factor 
$\Delta=(N+1)/2$, and the variance 
$\sigma^2$ of the components in the $\sigma$-subspace, 
related to the normalizing factor $a$ of the new features 
through $\sigma^2 = N a^2 / \Delta$. In our case, 
since $N=50$, $\Delta=25.5$ and $\sigma^2 = 1.960784 a^2$, 
that is $\sigma^2 = 1.960784$ for the non-normalized mapping 
and $\sigma^2 = 0.039216$, for the normalized one.

The values of $Q$, the fraction of squared student weights in the 
$\sigma$-subspace, and the teacher-student overlaps $R_u$ and 
$R_\sigma$, normalized within the corresponding sub-space, 
are represented on figures \ref{fig.Q*0} 
to \ref{fig.Q*05} as a function of $\alpha \equiv M/N$, 
using full and open symbols for the mappings ${\bbox \Phi}_n$ 
and ${\bbox \Phi}_{nn}$ respectively. Notice that the 
abscissas correspond to the fraction of training patterns 
per {\it input} space dimension. Error bars are smaller 
than the symbols' size. The lines are {\it not} fits, but the 
theoretical curves corresponding to the same classes of 
teachers as the experimental results. The excellent 
agreement with the experimental data is striking. 
Thus, the high order correlations of the 
features, neglected in the theoretical models, 
are indeed negligible. 

Fig. \ref{fig.Q*0} corresponds to a purely linear 
teacher ($Q^*=0$), i.e. to a quadratic SVM learning a 
rule linearly separable in input space. As in this case 
$R_\sigma=0$, only $R_u$ and $Q$ 
are represented. In the case of a purely quadratic rule, 
$Q^*=1$, represented on fig. \ref{fig.Q*1}, $R_u=0$. Notice 
that the corresponding overlaps, $R_u$ and $R_\sigma$, 
do not have a similar behaviour, as the latter increases 
much slower than the former, irrespective of the mapping. 
This happens because, as the number of quadratic components 
scales like $N \Delta$, a number of examples of the 
order of $N \Delta$ are needed to learn them. Indeed, 
$R_u$ reaches a value 
close to $1$ with $\alpha \sim O(1)$ while $R_\sigma$ 
needs $\alpha \sim O(\Delta)$ to reach similar values.

Fig. \ref{fig.Q*51/53} shows the results corresponding 
to the isotropic teacher, having $Q^*=Q^*_{iso} \equiv 
\Delta/(1+\Delta)$. For $\Delta=25.5$ we have $Q^*_{iso}=0.962$
A particular case of such a teacher has all its weight 
components of equal absolute value, i.e. $(w_i^*)^2 = 
1/{\tilde N}$, and was studied in~\cite{YO} and~\cite{DOS}. 
Finally, the results corresponding to a general 
rule, with $Q^*=0.5$, are shown in fig. \ref{fig.Q*05}. 
Notice that at fixed $\alpha$, $R_u$ decreases and 
$R_\sigma$ increases with $Q^*$ at a rate that 
depends on the mapping. These quantities 
determine the student's generalization error through 
the combination (\ref{eq.R}). The fact that they increase 
as a function of $\alpha$ with different speed is a signature of 
hierarchical learning. 

\begin{figure}
\centerline{\psfig{figure=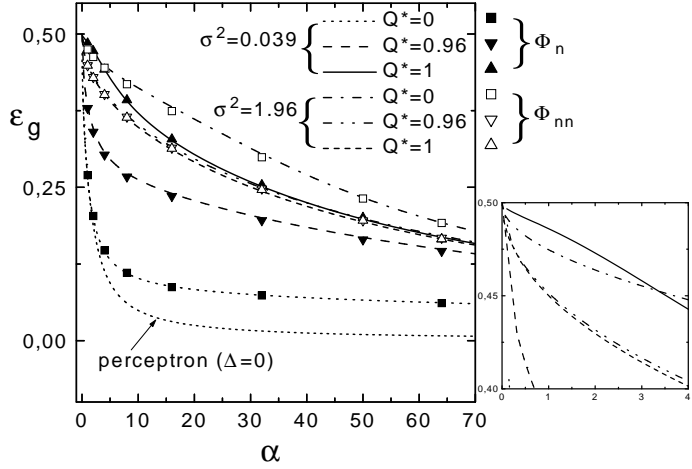,height= 7 cm}}
\caption{Learning curves of SVMs for different teacher 
rules $Q^*$. Definitions are the same as in figure 
{\protect {\ref{fig.Q*0}}}. The inset is an 
enlargement of the smalll $\alpha$ region.}
\label{fig.epsg}
\end{figure}

The generalization error $\epsilon_g$ corresponding 
to the different rules is plotted against 
$\alpha$ on fig. \ref{fig.epsg}, for both mappings. 
At any fixed $\alpha$, the performance obtained 
with the normalized mapping is better the smaller the value 
of $Q^*$. The non-normalized mapping shows the opposite 
trend: its performance for a purely linear teacher is 
extremely bad, but it improves for increasing values of 
$Q^*$ and slightly overrides that of the normalized 
mapping in the case of a purely quadratic teacher. 
These results reflect the competition on learning the 
anisotropically  distributed features. In the case of 
the normalized mapping, the $\sigma$-components are 
compressed ($\sigma^2=0.039$) with respect to the {\it u}-components, 
which have unit variance. This is advantageous whenever 
the linear components carry the most significant information, which 
is the case for $Q^* \ll 1$. When $Q^*=1$, the linear components 
only introduce noise that hinders the learning process. As the 
number of linear components is much smaller than the 
number of quadratic ones, their pernicious effect should be 
more conspicuous the smaller the value of $\Delta$. Conversely, 
the non-normalized mapping has $\sigma^2=1.96$, meaning that the 
compressed components are those of the {\it u}-subspace. Therefore, 
this mapping is better when most of the information is 
contained in the $\sigma$-subspace, which is the case for 
teachers with large $Q^*$ and, in particular, with $Q^*=1$.

\begin{figure}
\centerline{\psfig{figure=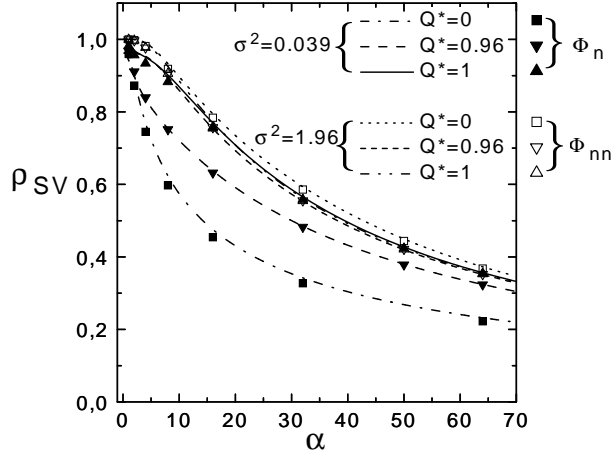,height= 7 cm}}
\caption{Fraction of learning patterns that belong to the subset of Support Vectors.}
\label{fig.rhoSV}
\end{figure} 

Finally, for the sake of completeness, the fraction of 
support vectors $\rho_{SV} \equiv M_{SV}/M$, where 
$M_{SV}$ is the number of training patterns with maximal 
stability, is represented on figure \ref{fig.rhoSV}. This 
fraction is an upper bound to the generalization error. 
Notice that these curves present qualitatively the same 
trends as $\epsilon_g$. Interestingly, $\rho_{SV}$ 
is smaller for the normalized mapping than for the 
non-normalized one for most of the rules. Since the student's 
weights can be expressed as a linear combination of SVs~\cite{Vapnik}, 
this result is of practical interest. 

\section{Discussion}
\label{sec.discussion}

In order to understand the results obtained in the 
previous section, we first analyze the relative 
behaviour of $R_u$ and $R_\sigma$, which can be 
deduced from equation (\ref{eq.sp4}). 
If $\Delta^*_\sigma \ll \Delta$, which is the case 
for sufficiently small $Q^*$, we get that $R_\sigma \ll R_u$. 
This means that the quadratic components are more difficult 
to learn than the linear ones. On the other 
hand, if the teacher lies mainly in the quadratic subspace, 
$\Delta^*_\sigma \gg \Delta$, and then $R_\sigma > R_u$. 
The crossover between these different behaviours occurs at 
$\Delta^*_\sigma = \Delta$, for which equation 
(\ref{eq.sp4}) gives $R_\sigma = R_u$. For $N=50$, 
which is the case in our simulations,  this 
arises for $Q^*_n=0.998$ or $Q^*_{nn}=0.929$, depending on whether 
we use the normalized or the non-normalized mapping. 
In the particular case of the isotropic teacher and the 
non-normalized mapping, $Q^* > Q^*_{nn}$, so that 
$R_\sigma > R_u$, as shown on figure \ref{fig.Q*51/53}.
These considerations alone are not sufficient to understand 
the behaviour of the generalization error, which depends on the 
weighted sum of $R_\sigma$ and $R_u$ (see equation 
(\ref{eq.R})).

The behaviour at small $\alpha$ is useful to understand the 
onset of hierarchical learning. A close inspection 
of equations (\ref{eq.sp1}-\ref{eq.sp4}) 
shows that in the limit $\alpha \rightarrow 0$, 
$x=\sigma^2$ and $Q \simeq \Delta \sigma^2 / (\Delta \sigma^2 +1)$ 
to leading order in $\alpha$. This results may be understood 
with the following simple argument: if 
there is only one training pattern, clearly it is a 
SV and the student's weight vector is proportional to 
it. As a typical example has $N$ components of 
unit length in the {\it u}-subspace and $N \Delta$ components 
of length $\sigma$ in the $\sigma$-subspace, we have 
$Q=N\Delta \sigma^2 / (N \Delta \sigma^2 +N)$. With the 
normalized mapping, $\lim_{\alpha \rightarrow 0} Q = 1/2$. 
In the case of the non normalized one $\lim_{\alpha \rightarrow 0} Q  
= (2 \Delta - 1)/2 \Delta$, which depends on the inflation factor 
of the SVM. In this limit, we obtain:

\begin{eqnarray}
\kappa_{max} &\simeq& 
\frac{1+\sigma^2 \Delta}{\sqrt{1+\sigma^4 \Delta}} \frac{1}{\sqrt{\alpha}}, \\
R_u &\simeq& \sqrt{\frac{2}{\pi}} \frac{1}{\sqrt{1+\Delta^*_\sigma}} 
\sqrt{\alpha}, \\
R_\sigma &\simeq& \sqrt{\frac{2}{\pi}} \, \sqrt{\frac{\Delta^*_\sigma}{1+\Delta^*_\sigma}} \, 
\sqrt{\frac{\alpha}{\Delta}}.
\end{eqnarray}

\noindent Therefore, $R \sim \sqrt{\alpha}$, like for the simple 
perceptron MMH~\cite{GG}, but with a prefactor that depends 
on the mapping and the teacher. 

In our model, we expect that hierarchical learning correspond to 
a fast increase of $R$ at small $\alpha$, mainly
dominated by the contribution of $R_u$. As in the limit 
$\alpha \rightarrow 0$,

\begin{equation}
R \simeq \frac{R_u + R_\sigma \sqrt{\sigma^4 \Delta \Delta^*_\sigma}}
{\sqrt{1+\sigma^4 \Delta} \sqrt{1+\Delta^*_\sigma}},
\end{equation}

\noindent we expect hierarchical learning if 
$\sigma^4 \Delta \ll 1$ and $\Delta^*_\sigma \lesssim 1$. 
The first condition establishes a constraint on the mapping, 
which is only satisfied by the normalized one. The second 
condition, that ensures that $R_\sigma < R_u$ holds, gives 
the range of teachers for which this hierarchical 
generalization takes place. Under these conditions, 
$R$ grows fast and the contribution 
of $R_\sigma$ is negligible because it is weighted 
by $\sqrt{\sigma^4 \Delta \Delta^*_\sigma}$. The effect 
of hierarchical learning is more important the smaller 
$\Delta^*_\sigma$. The most dramatic effect arises for $Q^*=0$, 
i.e. for a quadratic SVM learning a linearly separable rule.  

On the other hand, if 
$\sigma^4 \Delta \gg 1$, which is the case for the non 
normalized mapping, both $R_u$ and $R_\sigma$ contribute to 
$R$ with comparable weights. Notice that, if the normalized 
mapping is used, the condition 
$\Delta^*_\sigma \lesssim 1$ implies that $Q^* < Q^*_{iso} \equiv
\Delta/(1+\Delta)$, where $Q^*_{iso}$ corresponds to the isotropic 
teacher. A straightforward calculation shows that a fraction of $47.5 \%$ 
of teachers satisfies this constraint for $N=50$. In fact, the 
distribution of teachers as a function of $Q^*$ has its maximum 
at $Q^*_{iso}$. When $N \rightarrow \infty$, the distribution 
becomes $\delta(Q^* - Q^*_{iso})$, and $Q^*_{iso}$ tends to the 
median, meaning that in this limit, only about $50 \%$ of the 
teachers give raise to hierarchical learning when using 
the normalized mapping.

In the limit $\alpha \rightarrow \infty$, all the 
generalization error curves converge to the same asymptotic 
value as the simple perceptron MMH learning in the feature 
space, namely $\epsilon_g=0.500489 (1+\Delta)/ \alpha$, 
independently of $\sigma$ and $Q^*$. Thus, $\epsilon_g$ vanishes 
slower the larger the inflation factor $\Delta$.

Finally, it is worth to point out that for $\sigma = 1$, 
which would correspond to a normalizing factor $a=\sqrt{\Delta/N}$, 
the pattern distribution in feature space is isotropic. 
Irrespective of $Q^*$, the corresponding 
generalization error is exactly the same as that of a simple 
perceptron learning the MMH with isotropically 
distributed examples in feature space.

\begin{figure}
\centerline{\psfig{figure=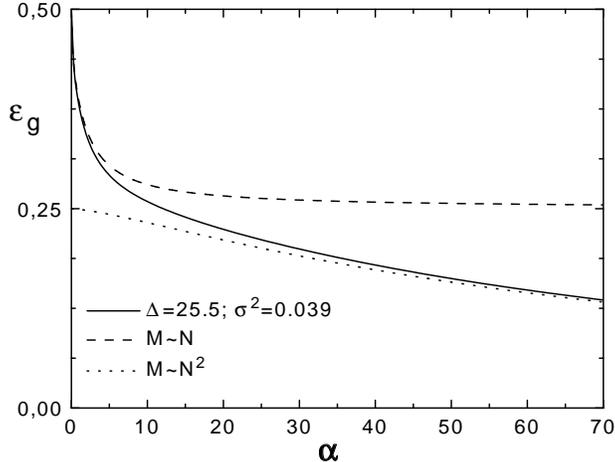,height= 7 cm}}
\caption{Generalization error of a SVM corresponding to 
different thermodynamic limits. See the text for the definition 
of $\alpha$ in each regime.}
\label{fig.thermolim}
\end{figure}

Since the inflation factor $\Delta$ of the SVM feature space 
in our approach is a free parameter, it does not diverge
in the thermodynamic limit $N \rightarrow \infty$ . As a 
consequence, $\epsilon_g$ does not present any stepwise 
behaviour, but just a crossover between a fast decrease 
at small $\alpha$ followed by a slower decrease regime at 
large $\alpha$. The results of Dietrich et al.~\cite{DOS} 
for the {\it normalized} mapping, that corresponds to $\sigma^2 
\Delta=1$ in our model, can be deduced by taking appropriately 
the limits before solving our saddle point equations. 
The regime where the number of training patterns $M= \alpha N$ 
scales with $N$, is straightforward. It is obtained by taking 
the limit $\sigma \rightarrow 0$ and $\Delta \rightarrow \infty$ keeping 
$\sigma^2 \Delta=1$ in our equations, with $\alpha$ finite. The 
regime where the number of training patterns 
$M= \alpha N$ scales with $N \Delta$, the number of quadratic 
features, obtained by keeping 
$\tilde \alpha \equiv \alpha/(1+ \Delta)$ finite whilst taking, 
here again, the limit $\sigma \rightarrow 0$, 
$\Delta \rightarrow \infty$ with $\sigma^2 \Delta=1$. 
The corresponding curves are represented on figure 
\ref{fig.thermolim} for the case of an isotropic teacher. 
In order to make the comparisons with our results at 
finite $\Delta$, the regime where $\tilde \alpha$ is 
finite is represented as a function of $\alpha = 
(1+\Delta) \tilde \alpha$ using the value of $\Delta$ 
corresponding to our numerical simulations, namely, 
$\Delta = 25.5$. In the same figure we represented the 
generalization error $\epsilon_g=(1/\pi) \arccos(R)$ 
where $R$, given by eq. (\ref{eq.R}), is obtained after 
solving the saddle point equations with parameter 
values $\sigma^2=0.039$ and $\Delta=25.5$.

These results, obtained for quadratic SVMs, are easily 
generalizable to higher order polynomial SVMs. The 
corresponding saddle point equations are cumbersome, 
and will not be given here. We expect a cascade of 
hierarchical generalization behaviour, 
in which successively more and more compressed features are 
learned. This may be understood by considering the 
set of saddle point equations that generalize 
equation (\ref{eq.sp4}). These equations relate the 
teacher-student overlaps in the successive subspaces. 
The sequence of different feature subspaces generalized by the 
SVM depends on the relative complexity of the teacher and 
the student. This is contained in the factors 
$\Delta^*_{\sigma_m}/\Delta_m$ corresponding to the 
$m^{th}$ subspace, that appear in the set of equations 
that generalize eq. (\ref{eq.sp4}).

\section{Conclusion}
\label{sec:conclusion}

We introduced a model that clarifies some aspects of the 
generalization properties of polynomial Support Vector 
Machines (SVMs) in high dimensional feature spaces. To 
this end, we focused on quadratic SVMs. The quadratic 
features, which are the pairwise products of input 
components, may be scaled by a {\it normalizing factor}. 
Depending on its value, the generalization 
error presents very different behaviours in the 
thermodynamic limit~\cite{DOS,BG}. 

In fact, a finite size SVM may be caracterized by two 
parameters: $\Delta$ and $\sigma$. The 
{\it inflation factor} $\Delta$ is the 
ratio between the quadratic and the linear 
features dimensions. Thus, it is proportional 
to the input space dimension $N$. The {\it variance} 
$\sigma$ of the quadratic features is related to 
the corresponding normalizing factor. Usually, either
$\sigma \sim 1/\sqrt{N}$ (normalized mapping) or 
$\sigma \sim 1$ (non normalized mapping). 
In previous studies, not only the input space dimension 
diverges in the thermodynamic limit $N \rightarrow \infty$, 
but also $\Delta$ and $\sigma$ are correspondingly scaled.

In our model, neither the proportion of quadratic features 
$\Delta$ nor their variance $\sigma$ are necessarily related 
to the input space dimension $N$. They are considered as 
parameters caracterizing the SVMs. Since we keep them constant 
when taking the thermodynamic limit, we can study the learning 
properties of actual SVMs with finite inflation ratios and 
normalizing factors, as a function of $\alpha \equiv M/N$, 
where $M$ is the number of training examples. Our theoretical 
results were obtained neglecting the correlations among the 
quadratic features. The agreement between our computer 
experiments with actual SVMs and the theoretical 
predictions is excellent. The effect of the correlations 
does not seem to be important, as there is almost no 
difference between the theoretical curves and the numerical 
results.

We find that the generalization error $\epsilon_g$ 
depends on the type of rule to be inferred through 
$Q^*$, the (normalized) sum of the teacher's squared weight 
components in the quadratic subspace. If 
$Q^*$ is small enough, the quadratic components 
need more patterns to be learned than the linear 
ones. However, only if the quadratic features 
are normalized, $\epsilon_g$ is dominated by 
the high rate learning of the linear components
at small $\alpha$. Then, on increasing $\alpha$, 
there is a crossover to a regime 
where the decrease of $\epsilon_g$ becomes much slower. 
The crossover between these two behaviours is smoother 
for larger values of $Q^*$, and this effect of 
hierarchical learning disappears for large enough $Q^*$. 
On the other hand, if the features are not normalized, 
the contributions of both the linear and the quadratic 
components to $\epsilon_g$ are of the same order, and 
there is no hierarchical learning at all. 

In the case of the normalized mapping, if the limits 
$\Delta \sim N \rightarrow \infty$ 
and $\sigma^2 \sim 1/N \rightarrow 0$
are taken together with the thermodynamic limit, the 
hierarchical learning effect gives raise to the two different 
regimes, corresponding to $M \sim N$ or $M \sim N^2$, described previously~\cite{YO,DOS}.
  
It is worth to point out that if the rule to be learned 
allows for hierarchical learning, the generalization 
error of the normalized mapping is much smaller than 
that of the non normalized one. In fact, 
the teachers corresponding to such rules are those 
with $Q^* \lesssim Q^*_{iso}$, where $Q^*_{iso}$ 
corresponds to the isotropic teacher, the one 
having all its weights components equal. For the others, both 
the normalized mapping and the non normalized one 
present similar performances. If the weights of the 
teacher are selected at random on a hypersphere 
in feature space, the most probable 
teachers have precisely $Q^*=Q^*_{iso}$, and the fraction 
of teachers with $Q^* \leq Q^*_{iso}$ represent of the order of 
$50\%$ of the inferable rules. Thus, from a practical 
point of view, without having any prior knowledge 
about the rule underlying a set of examples, the 
normalized mapping should be preferred.
 
\section*{Acknowledgements}

It is a pleasure to thank Arnaud Buhot for a careful 
reading of the manuscript, and Alex Smola for 
providing us the Quadratic 
Optimizer for Pattern Recognition program~\cite{AS}. 
The experimental results were obtained with the Cray-T3E 
computer of the CEA (project 532/1999).

SR-G acknowledges economic support from the  EU-research contract 
ARG/B7-3011/94/97. 

MBG is member of the CNRS.

\end{document}